\documentclass{elsart}

\usepackage{graphicx}
\usepackage{amssymb}
\usepackage{epsfig}
\usepackage{subfigure}

\usepackage[english]{babel}
\usepackage{latexsym}
\usepackage{amsfonts}
\usepackage{amsmath}
\usepackage{longtable}

\def\be{\begin{equation}}
\def\ee{\end{equation}}

\begin{document}
\begin{frontmatter}

\title{The canonical partition function for relativistic hadron gases}

\author[hei]{F.~Beutler},
\author[gsi]{A.~Andronic},
\author[gsi,emmi,tud,fias]{P.~Braun-Munzinger},
\author[wro,emmi,tud]{K.~Redlich},
\author[hei]{J.~Stachel}

\address[hei]{Physikalisches Institut der Universit\"at Heidelberg,
D-69120 Heidelberg, Germany}
\address[gsi]{GSI Helmholtzzentrum f\"ur Schwerionenforschung, D-64291
Darmstadt, Germany}
\address[emmi]{ExtreMe Matter Institute EMMI, GSI, D-64291 Darmstadt, Germany}
\address[tud]{Technical University Darmstadt, D-64289 Darmstadt, Germany}
\address[fias]{Frankfurt Institute for Advanced Studies, J.W. Goethe University,
D-60438 Frankfurt, Germany}
\address[wro]{Institute of Theoretical Physics, University of Wroc\l aw,
PL-50204 Wroc\l aw, Poland}

\begin{abstract}
Particle production in high-energy collisions is often addressed within
the framework of the thermal (statistical) model.
We present a method to calculate  the canonical partition function for the 
hadron resonance gas with exact conservation of  the baryon number, 
strangeness, electric charge, charmness and bottomness.
We derive an analytical expression for the partition function which is 
represented as series of Bessel functions. Our results can be used directly 
to analyze particle production yields in elementary and in heavy ion 
collisions. 
We also quantify the importance of quantum statistics in the calculations 
of the light particle multiplicities in the canonical thermal model of the
hadron resonance gas.

\end{abstract}


\end{frontmatter}

\section{Introduction}

The statistical model of the hadron resonance gas was shown to be very  successful
in addressing particle production in high-energy collisions 
\cite{review,aa05,becgaz,man08,letessier05,aa08,Becattini:1995if,Cleymans:1998yb,Andronic:2008ev,becattini08,Cleymans:2006xj,Kraus:2007bi}.
In central nucleus-nucleus collisions the formulation of conservation of quantum numbers 
(sometimes generically referred to as charges) 
in the grand canonical (GC) ensemble is well suited 
\cite{aa05,becgaz,man08,letessier05,aa08}. However, for small systems like 
$e^+e^-$ or pp as well as for peripheral nucleus-nucleus collisions only 
the canonical (C) ensemble gives the correct description of hadron yields 
\cite{review,Becattini:1995if,Cleymans:1998yb,Andronic:2008ev,becattini08,Cleymans:2006xj,Kraus:2007bi}. 
In the GC formulation the conservation of charges is implemented  on
the average and is controlled by the appropriate chemical potentials whereas 
in the C-ensemble charges are conserved exactly \cite{review}.

The exact treatment of quantum numbers in statistical mechanics is well
established \cite{Hagedorn:1970gh,Shuryak:1973pv,Redlich:1979bf,Rafelski:1980gk,Turko:1981nr,Muller:1982gd,Hagedorn:1984uy}.
It is in general obtained by projection of the partition function on the desired
values of the conserved charges by using group theoretical methods
\cite{Redlich:1979bf,Turko:1981nr,Muller:1982gd,Turko:2000if,Redlich:2003dw}.
For studies of hadron production one needs in general to account
for the exact conservation of five quantum numbers:
 the baryon number $N$, strangeness $S$, electric
charge $Q$, charmness $C$, and bottomness $B$. The conservation is a direct consequence of
 the $G=U_N(1)\otimes ..\otimes U_B(1)$ invariance of strong interactions.

The implementation of an exact conservation of  quantum numbers in the 
canonical partition function for the hadron resonance gas requires integration over 
the symmetry group \cite{review,Redlich:1979bf}.
This in general leads to numerical problems due to the oscillatory behavior
of the integrand. To avoid numerical  problems, methods were developed  
to express the partition function as  series of Bessel  functions which 
is well suited for numerical implementations
\cite{Cleymans:1990mn,Cleymans:1991yu,Cleymans:1997ib,BraunMunzinger:2001as}.
An explicit  analytical expression for the canonical partition function using 
the above methods was obtained up to now only for three conserved  
quantum numbers i.e. for $N$, $S$ and $Q$ and employing Boltzmann statistics
\cite{Cleymans:1997ib}. 

In the application of the statistical model to hadron production, e.g. 
in  $e^+e^-$ annihilation, one must employ the formulation of the canonical 
partition function which accounts for the conservation of all five quantum 
numbers including charmness and botomness since hadrons carrying
 heavy quarks contribute sizably to the overall hadron yields 
\cite{Becattini:1995if,Andronic:2008ev,becattini08,pdg}.
In addition, for $e^+e^-$  collisions,  the data are measured with high 
precision; see \cite{pdg} for a recent summary. Consequently,  in the 
statistical model analysis the approximation of particle momentum  
distributions using Boltzmann statistics  is  not sufficient 
\cite{Andronic:2008ev,Redlich:2009xx}.

The main scope of this paper is the extension of the previous results for 
the canonical partition function for a hadron gas \cite{Cleymans:1997ib} 
to include the exact conservation of all quantum numbers carried 
by the light and heavy quarks \cite{fb_dipl}. 
We will formulate the canonical partition 
function as series of the modified Bessel functions. Our expression  
is numerically stable and can be used to quantify thermodynamics independently 
of the values of the thermal parameters and the initial quantum numbers. 
Our result accounts for the quantum statistics for bosons.  
For all fermions the masses are significantly larger than the temperature. 
Consequently, the implementation of Boltzmann statistics for fermions 
is a very good approximation. An extension of our results to quantum statistics
for all particles including fermions would however be quite straightforward.

As model implementation of the canonical partition function we illustrate 
its application to the  calculations of hadron yields. We  quantify 
the  quantum statistics  effects for the multiplicity of pions and kaons.
We also discuss  deviations from the exact quantum statistics results 
when the Boltzmann approximation is used for different values of the thermal 
parameters.  
The hadron resonance gas we are working with is containing all known
hadrons, including the multi-strange hyperons up to $S=\pm 3$ and all charmed
and bottom hadrons, as listed in the latest compilation by the Particle Data 
Group \cite{pdg}.

The paper is organized as follows: in the next Section we derive the analytical 
expression for the canonical partition function that accounts for the 
conservation of five quantum numbers. In Section 3 we present a particular 
numerical implementation to calculate the pion and kaon multiplicities and 
discuss the role of quantum statistics. We summarize our results in Section 4.

\section{The canonical partition function}

The appropriate tool to deal in a statistical mechanics framework
with a system  of quantum numbers  $\vec X=(N,S,Q,C,B)$ related with the 
$G=U_N(1)\otimes ..\otimes U_B(1)$ symmetry group
 is the canonical partition function \cite{Turko:1981nr} ($\hbar=c=1$):
\begin{equation}
\label{eq:partition6}
\begin{split}
\mathcal{Z}_{N,S,Q,C,B}(\vec{X}) &= \frac{1}{(2\pi)^5}\int d^5\vec{\phi}\; e^{i\vec{X}\vec{\phi}}\\
&\quad \exp\left\{\sum_j\frac{g_jV}{(2\pi)^3}\int d^3p \ln(1\pm e^{-\frac{\sqrt{\vec{p}^2+m_j^2}}{T}-i\vec{x}_j\vec{\phi}})^{\pm1}\right\}
\end{split}
\end{equation}
where the  vector $\vec{X}$=$(N,S,Q,C,B)$ characterizes  the
initial quantum numbers of a system  related with  each of the $U(1)$ symmetry 
groups and the $\pm$ sign refers to fermions and bosons. 
The exact quantum number conservation is implemented by the  integration 
over the group $G$ with  $\vec{\phi} = (\phi_N,\phi_S,\phi_Q,\phi_C,\phi_B)$  
being an element of $G$.
The vector $\vec{x}_j=(N_j,S_j,Q_j,C_j,B_j)$ describes the quantum
numbers of a particle $j$ with mass $m_j$ and  the spin-isospin degeneracy 
factor $g_j$.
The sum in the exponential is taken over all particles and resonances which 
carry  quantum numbers related with an  internal symmetry $G$.

The partition function  (\ref{eq:partition6}) depends only on two parameters, the
 temperature $T$ and the volume $V$ of the system.
However, the integral representation of this partition function is very 
inconvenient for numerical  implementations. This is particularly the case 
for non-vanishing initial quantum numbers due to the oscillatory nature of 
the integrand. An additional complication appears due to quantum 
statistics effects. Clearly, one could simplify the problem and  employ the
Boltzmann approximation by  keeping only the first term of the series expansion 
of the $\ln(1\pm x)^{\pm1}$ in the exponential. However, such an approximation 
is only  valid if the particle mass is significantly larger than the temperature.
As the values of temperatures extracted from the fits of hadron abundances 
obtained  in heavy ion and elementary collisions
\cite{review,aa05,Becattini:1995if,Andronic:2008ev,becattini08} are
close to the pion mass, the Boltzmann approximation will give rise to
deviations  from the correct quantum statistics values  in particular for 
light particles like pions or kaons.
To calculate the partition function accurately (using quantum statistics)
one needs in general to include the whole series of the expansion of the logarithm
\begin{equation}
\ln(1\pm x)^{\pm 1} = \sum^{\infty}_{k=1}(\pm 1)^{k+1}\frac{x^{k}}{k}
\label{eq:partition19.2}
\end{equation}
in the partition function for all particle species. However,
as the fermions are heavy, the error caused by the Boltzmann approximation
in this case is very small. Therefore we will use the whole series of 
the logarithm only for light bosons (mesons without charm or bottom quarks) 
and apply the usual Boltzmann approximation to fermions and heavy bosons 
(bosons containing charm or  bottom quarks).
With this approximation equation (\ref{eq:partition6}) becomes:
\begin{equation}\label{eq:partition19.3}
\begin{split}
\mathcal{Z}_{N,S,Q,C,B}(\vec{X}) &= \frac{1}{(2\pi)^5}\int d^5\vec{\phi}\; e^{i\vec{X}\vec{\phi}}\exp\left\{\sum_j z^1_je^{-i\vec{x}_j\vec{\phi}} + \sum_{b}\sum_{k=2}^{\infty}z^k_be^{-ik\vec{x}_b\vec{\phi}}\right\}
\end{split}
\end{equation}
with the particle partition function
\begin{equation}
z^k_j = \frac{g_jV}{k(2\pi)^3}\int d^3p\;
e^{-\frac{\sqrt{\vec{p}^2+m_j^2}}{T}k},
\label{eq:partition19.4}
\end{equation}
where  $j$ runs over all particles, while $b$ runs only over the light bosons.

 For charm and bottom hadrons, because of their large masses  the term   
$\sum_{j_{c/b}}z^1_{j_{c/b}}$ is much less\footnote{The particle partition function 
$z^1_{j_{c/b}}$ is $\mathcal{O}(10^{-4})$ for charm particles and 
$\mathcal{O}(10^{-13})$ for bottom particles} than 1. 
Consequently, we can use the following approximation \cite{Becattini:1995if}:
\begin{equation}
\exp\left\{\sum_{j_{c/b}}z^1_{j_{c/b}}e^{-i\vec{x}_j\vec{\phi}}\right\} \simeq 1 + \sum_{j_{c/b}}z^1_{j_{c/b}}e^{-i\vec{x}_{j_{c/b}}\vec{\phi}}
\label{eq:partition9}
\end{equation}
Inserting now Eq. (\ref{eq:partition9}) into Eq.  (\ref{eq:partition19.3}) we 
obtain:
\begin{equation}\label{eq:partition10}
\begin{split}
\mathcal{Z}_{N,S,Q,C,B}(\vec{X}) & \approx \frac{1}{(2\pi)^{5}}\int d^{5}\vec{\phi}\; e^{i\vec{X}\vec{\phi}} e^{f(\vec{\phi})}\\
&\quad + \sum_{j_{c}}z^1_{j_{c}}\frac{1}{(2\pi)^{5}}\int d^{5}\vec{\phi}\; e^{i(\vec{X}-\vec{x}_{j_{c}})\vec{\phi}} e^{f(\vec{\phi})}\\
&\quad + \sum_{j_{b}\;\;\&\atop C_{j_b} = 0}z^1_{j_{b}}\frac{1}{(2\pi)^{5}}\int d^{5}\vec{\phi}\; e^{i(\vec{X}-\vec{x}_{j_{b}})\vec{\phi}} e^{f(\vec{\phi})}\\
&\quad + \sum_{j_{c}}\sum_{j_{b}\;\;\&\atop C_{j_b} = 0}z^1_{j_{c}}z^1_{j_{b}}\frac{1}{(2\pi)^{5}}\int d^{5}\vec{\phi}\; e^{i(\vec{X}-\vec{x}_{j_{c}}-\vec{x}_{j_{b}})\vec{\phi}} e^{f(\vec{\phi})}
\end{split}
\end{equation}
with
\begin{equation}\label{eq:fphi}
f(\vec{\phi}) = \sum_jz^1_je^{-i\vec{x}_j\vec{\phi}} + \sum_{b}\sum_{k=2}^{\infty}z^k_be^{-ik\vec{x}_b\vec{\phi}}.
\end{equation}
The index  $j$ in Eq. (\ref{eq:fphi}) runs  over all hadrons except 
those  which carry   heavy flavors whereas in Eq. (\ref{eq:partition10}) 
$j_c$ runs over all charm  and $j_b$  over all bottom hadrons.

From Eq.(\ref{eq:partition10}) it is transparent  that the  integrals over 
$\phi_C$ and $\phi_B$ related with the charm and bottom quantum numbers 
contribute  to the partition function as Kronecker delta functions.
Consequently,
\begin{equation}\label{eq:partition11}
\begin{split}
\mathcal{Z}_{N,S,Q,C,B}(\vec{X}) & \approx \frac{1}{(2\pi)^{3}}\int d^{3}\vec{\phi}\; e^{i\vec{X}\vec{\phi}} e^{f(\vec{\phi})} \delta_{C,0}\delta_{B,0}\\
&\quad + \sum_{j_{c}}z^1_{j_{c}}\frac{1}{(2\pi)^{3}}\int d^{3}\vec{\phi}\; e^{i(\vec{X}-\vec{x}_{j_{c}})\vec{\phi}} e^{f(\vec{\phi})} \delta_{C,C_{j_{c}}}\delta_{B,B_{j_c}}\\
&\quad + \sum_{j_{b} \;\;\&\atop C_{j_b} = 0}z^1_{j_{b}}\frac{1}{(2\pi)^{3}}\int d^{3}\vec{\phi}\; e^{i(\vec{X}-\vec{x}_{j_{b}})\vec{\phi}} e^{f(\vec{\phi})} \delta_{C,0}\delta_{B,B_{j_{b}}}\\
&\quad + \sum_{j_{c}}\sum_{j_{b}\;\;\&\atop C_{j_b} = 0}z^1_{j_{c}}z^1_{j_{b}}\frac{1}{(2\pi)^{3}}\int d^{3}\vec{\phi}\; e^{i(\vec{X}-\vec{x}_{j_{c}}-\vec{x}_{j_{b}})\vec{\phi}} e^{f(\vec{\phi})} \delta_{C,C_{j_{c}}}\delta_{B,B_{j_{c}}+B_{j_b}}
\end{split}
\end{equation}
where $\vec{X}$ and $\vec{x}_j$ are now three-dimensional vectors composed of 
the baryon number, the strangeness and the electric charge, while the charmness
$C$ and bottomness $B$ appear only through  the Kronecker functions.

With the approximation  (\ref{eq:partition9}) for charm and bottom contributions 
to the generating functional  we have reduced the five dimensional integrations 
to three dimensional ones in the canonical partition function. The integrals 
to compute are of  the following generic form:
\begin{equation}\label{eq:partition12}
\begin{split}
\mathcal{I}_{N,S,Q} &= \frac{1}{(2\pi)^3}\int^{2\pi}_0 d^3\vec{\phi}\;e^{i\vec{X}\vec{\phi}}e^{f(\vec{\phi})} 
\end{split}
\end{equation}
and correspond to the canonical  partition function with the conservation of 
three quantum numbers that  accounts for the quantum statistics of  bosons.

\begin{table}[htb]
\caption{The combinations of quantum numbers ($N$=baryon number, $S$=strangeness,
$Q$=electric charge) for hadrons and the corresponding notation for the 
sum of particle partition functions for each hadron class, $Z_{hadr}$. 
The correspondence of $Z_{hadr}$ to the index $n_j$ in 
Eq.~\ref{eq:partition22} (see text) is also given.}
\label{tabl:hadr}
\begin{center}
\begin{tabular}{ccccc}
  \hline
  \hline
  \multicolumn{3}{c}{Quantum numbers} & $Z_{hadr}$ & Index $n_j$\\
  \hline
  $N$=0 & $S$=0 & $Q$=0 & $Z_0$ & - \\
  $N$=0 & $S$=1 & $Q$=0 & $Z_{K^0}$ & - \\ 
  $N$=1 & $S$=0 & $Q$=0 & $Z_n$ & -\\
  $N$=0 & $S$=0 & $Q$=1 & $Z_{\pi^{\pm}}$ & -\\ 
  $N$=1 & $S$=0 & $Q$=1 & $Z_{p}$ & $n_1$ \\
  $N$=1 & $S$=0 & $Q$=-1 & $Z_{\Delta^{\mp}}$ & $n_2$\\
  $N$=1 & $S$=0 & $Q$=2 & $Z_{\Delta^{++}}$ & $n_3$\\
  $N$=0 & $S$=1 & $Q$=1 & $Z_{K^{\pm}}$ &  $n_4$ \\ 
  $N$=1 & $S$=-1 & $Q$=0 & $Z_{\Lambda}$ & $n_5$\\
  $N$=1 & $S$=-1 & $Q$=1 & $Z_{\Sigma^+}$ & $n_6$\\
  $N$=1 & $S$=-1 & $Q$=-1 & $Z_{\Sigma^-}$ & $n_7$\\
  $N$=1 & $S$=-2 & $Q$=0 & $Z_{\Xi^0}$ & $n_8$ \\
  $N$=1 & $S$=-2 & $Q$=-1 & $Z_{\Xi^{\mp}}$ & $n_9$\\
  $N$=1 & $S$=-3 & $Q$=-1 & $Z_{\Omega^{\mp}}$ & $n_{10}$\\
  \hline
  \hline
\end{tabular}
\end{center}
\end{table}

The integral representation of $\mathcal{I}_{N,S,Q}$  is not convenient
for numerical analysis as the integrand is a strongly oscillatory function, 
particularly for large initial quantum numbers $N$, $S$ or $Q$. 
Following the methods described in Refs. \cite{Cleymans:1990mn,Cleymans:1991yu} 
and \cite{Cleymans:1997ib} we  express $\mathcal{I}_{N,S,Q}$ as a series  
of  Bessel-functions.
First, we observe that, in  the argument of the exponential function in
Eq. (\ref{eq:partition12}),  the  particles  appear pairwise with their  
anti-particles. Second, the contributions of all particles in the sum  
can be grouped into 14 categories defined by their
quantum numbers, see Table \ref{tabl:hadr} for definitions. 
For instance $Z_{K^0}$ describes
the sum over all  partition functions $z^1_j$ of particles $j$ with $N$=0,
$S$=1 and $Q$=0. Consequently,
we  rewrite the integral   (\ref{eq:partition12}) as follows:
\begin{equation}\label{eq:partition20}
\begin{split}
\mathcal{I}_{N,S,Q} &= \exp(Z_0) 
\frac{1}{2\pi}\int^{2\pi}_0 d\phi_N\; e^{iN\phi_N}\exp[Z_n(e^{i\phi_N} + e^{-i\phi_N})]\\
&\quad\frac{1}{2\pi}\int^{2\pi}_0 d\phi_S\; e^{iS\phi_S} \exp[Z_{K^0}(e^{i\phi_S} + e^{-i\phi_S})]\\
&\quad\frac{1}{2\pi}\int^{2\pi}_0 d\phi_Q\; e^{iQ\phi_Q}\exp[Z_{\pi^{\pm}}(e^{i\phi_Q} + e^{-i\phi_Q})]\\
&\quad\exp[Z_{p}(e^{i(\phi_N+\phi_Q)} + e^{-i(\phi_N+\phi_Q)})]\\
&\quad\exp[Z_{\Delta^{\mp}}(e^{i(\phi_N-\phi_Q)} + e^{-i(\phi_N-\phi_Q)})]\\
&\quad\exp[Z_{\Delta^{++}}(e^{i(\phi_N+2\phi_Q)} + e^{-i(\phi_N+2\phi_Q)})]\\
&\quad\exp[Z_{K^{\pm}}(e^{i(\phi_S+\phi_Q)} + e^{-i(\phi_S+\phi_Q)})]\\
&\quad\exp[Z_{\Lambda}(e^{i(\phi_N-\phi_S)} + e^{-i(\phi_N-\phi_S)})]\\
&\quad\exp[Z_{\Sigma^+}(e^{i(\phi_N-\phi_S+\phi_Q)} + e^{-i(\phi_N-\phi_S+\phi_Q)})]\\
&\quad\exp[Z_{\Sigma^-}(e^{i(\phi_N-\phi_S-\phi_Q)} + e^{-i(\phi_N-\phi_S-\phi_Q)})]\\
&\quad\exp[Z_{\Xi^0}(e^{i(\phi_N-2\phi_S)} + e^{-i(\phi_B-2\phi_S)})]\\
&\quad\exp[Z_{\Xi^{\mp}}(e^{i(\phi_N-2\phi_S-\phi_Q)} + e^{-i(\phi_N-2\phi_S-\phi_Q)})]\\
&\quad\exp[Z_{\Omega^{\mp}}(e^{i(\phi_N-3\phi_S-\phi_Q)} + e^{-i(\phi_N-3\phi_S-\phi_Q)})]\\
&\quad\exp\left[\sum^{\infty}_{k=2}Z^k_{\pi^{\pm}}(e^{ik\phi_Q} + e^{-ik\phi_Q})\right]\\
&\quad\exp\left[\sum^{\infty}_{h=2}Z^h_{K^{0}}(e^{ih\phi_S} + e^{-ih\phi_S})\right]\\
&\quad\exp\left[\sum^{\infty}_{l=2}Z^l_{K^{\pm}}(e^{il(\phi_S+\phi_Q)} + e^{-il(\phi_S+\phi_Q)})\right].
\end{split}
\end{equation}

Applying the relation:
\begin{equation}
\exp\left[\frac{x}{2}\left(t+\frac{1}{t}\right)\right] = \sum^{\infty}_{m=-\infty}t^mI_m(x)
\label{eq:partition13}
\end{equation}
and the integral representation of the Bessel-function of order $h$:
\begin{equation}
I_h(x) = \frac{1}{2\pi}\int^{2\pi}_0 d\phi \exp(x\cos\phi)\exp(-ih\phi),
\label{eq:partition16.2}
\end{equation}
the group integrals in  Eq. (\ref{eq:partition20}) can be done explicitly, 
yielding:
\begin{equation}\label{eq:partition22}
\begin{split}
\mathcal{I}_{N,S,Q} & = \exp(Z_0) 
\prod^{10}_{j=1}\left[\sum^{\infty}_{n_j={-\infty}}I_{n_j}(2Z_{hadr})\right]
\prod^{\infty}_{k=2}\left[\sum^{\infty}_{m_k=-\infty}I_{m_k}(2Z^{k}_{\pi^{\pm}})\right]
\prod^{\infty}_{h=2}\left[\sum^{\infty}_{m_h=-\infty}I_{m_h}(2Z^{h}_{K^0})\right] \\
&\quad
\prod^{\infty}_{l=2}\left[\sum^{\infty}_{m_l=-\infty}I_{m_l}(2Z^{l}_{K^{\pm}})\right]
I_{-\nu_1}(2Z_{K^0}) I_{-\nu_2}(2Z_n) I_{-\nu_3}(2Z_{\pi^{\pm}}),
\end{split}
\refstepcounter{equation}\tag{\theequation}
\end{equation}
with
\begin{align}
\nu_1 &= Q+n_1-n_2+2n_3+n_4+n_6-n_7-n_9-n_{10}+\sum km_k+\sum lm_l\nonumber\\
\nu_2 &= S+n_4-n_5-n_6-n_7-2n_8-2n_9-3n_{10}+\sum hm_h+\sum lm_l\nonumber\\
\nu_3 &= N+n_1+n_2+n_3+n_5+n_6+n_7+n_8+n_9+n_{10},\nonumber
\end{align}
and with the specific $Z_{hadr}$ pertaining to each index $n_j$ according to 
Table~\ref{tabl:hadr}.
The above equation  is a generalization of that for a canonical partition function 
with exact conservation of baryon number, electric charge and 
strangeness  \cite{Cleymans:1997ib} to the  case  where  
quantum statistics for bosons is explicitly included.

To obtain  the partition function conserving five quantum numbers we use
Eq.~(\ref{eq:partition12}) to transform Eq.~(\ref {eq:partition11}) into:
\begin{equation}\label{eq:partition18}
\begin{split}
\mathcal{Z}_{N,S,Q,C,B}(\vec{X}) & \approx \mathcal{I}_{N,S,Q}\;\delta_{C,0}\delta_{B,0}\\
&\quad +  \sum_{j_{c}}z^1_{j_{c}}\mathcal{I}_{N+N_{j_c},S+S_{j_c},Q+Q_{j_c}} \delta_{C,C_{j_{c}}}\delta_{B,0}\\
&\quad +  \sum_{j_{b} \;\;\&\atop C_{j_b} = 0}z^1_{j_{b}}\mathcal{I}_{N+N_{j_b},S+S_{j_b},Q+Q_{j_b}} \delta_{C,0}\delta_{B,B_{j_{b}}}\\
&\quad +  \sum_{j_{c}}\sum_{j_{b}\;\;\&\atop C_{j_b} = 0}z^1_{j_{c}}z^1_{j_{b}}\mathcal{I}_{N+N_{j_c}+N_{j_b},S+S_{j_c}+S_{j_b},Q+Q_{j_c}+Q_{j_b}} \delta_{C,C_{j_{c}}}\delta_{B,B_{j_{c}}+B_{j_b}}.
\end{split}
\end{equation}
This expression, together with  (\ref{eq:partition22}) is our final result 
for the partition function that accounts for exact  conservation of 
baryon number, electric charge,  strangeness, charmness and bottomness. 
The  partition function (\ref{eq:partition18}), contrary to its integral 
representation (\ref{eq:partition6}), is free from oscillations and is 
numerically stable  independent of the values of the thermal parameters or  
the values of initial quantum numbers.

The partition function (\ref{eq:partition18}) can be used to describe 
thermodynamical properties of the hadron resonance gas  under  constraints  
of the exact conservation of all relevant quantum numbers.
In particular, from Eq. (\ref{eq:partition18}) we obtain the multiplicity 
$\langle n_j\rangle$  for hadron species $j$ by introducing a
fugacity parameter $\lambda_j$ which multiplies the particle partition 
function $z_j$ and by differentiating:
\begin{equation}
\left.\langle n_j\rangle = \frac{\partial\ln \mathcal{Z}_{N,S,Q,C,B}}{\partial
\lambda_j}\right|_{\lambda_j=1}. \label{eq:partition2}
\end{equation}
 For bosons, e.g. for $\pi^{\pm}$, one obtains from Eqs. (\ref{eq:partition18}) 
and (\ref{eq:partition2}),
\begin{align}\label{last}
\langle n_{\pi^{\pm}}\rangle &= \left.\frac{\partial\ln \mathcal{Z}}{\partial \lambda_{\pi^{\pm}}}\right|_{\lambda_{\pi^{\pm}}=1}\notag\\
&= \sum^{\infty}_{k=1}kz^k_{\pi^{\pm}}\frac{\mathcal{Z}_{N,S,Q,C,B}(\vec{X}-k\vec{x}_{\pi^{\pm}})}{\mathcal{Z}_{N,S,Q,C,B}(\vec{X})}
\end{align}
whereas for fermions there is only one term contributing ($k$=1) because 
fermions are well approximated by Boltzmann statistics.

\section{Numerical results}

We have applied the above obtained analytical expression for the canonical partition 
function (\ref{eq:partition18}) to quantify particle production in $e^+e^-$ 
annihilations at LEP energies  \cite{Andronic:2008ev,aa2009}.
We will not repeat the discussion given there but focus, in the following on an 
illustration of the importance of quantum statistics in the 
calculation of  multiplicities  of light bosons.

\begin{figure}[htb]
\begin{center}
\includegraphics[width=0.65\textwidth]{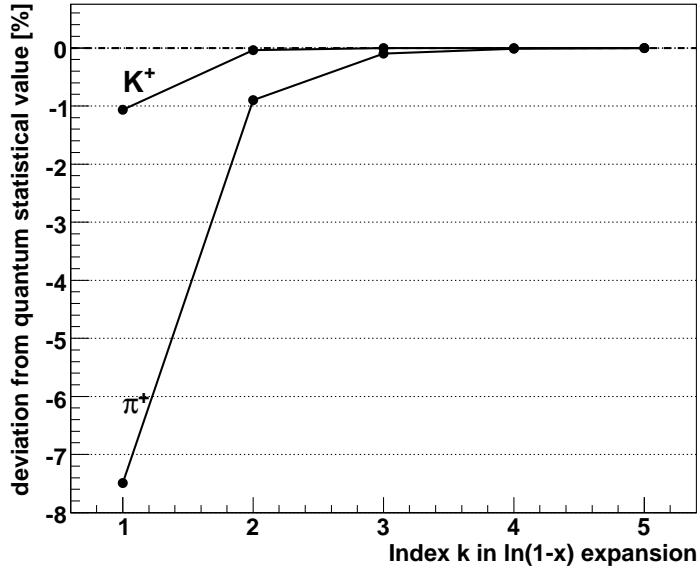}
\caption{The deviation of the $\pi^+$ and $K^+$ yields from the quantum
statistical value caused by the use of Boltzmann statistics as a function of the
index $k$ ($h$, $l$) (see Eq.~\ref{eq:partition19.2}).}
\label{fig:dev1}
\end{center}
\end{figure}

Fig.~\ref{fig:dev1} shows relative deviations of pion and kaon
multiplicities from their quantum statistics values with
increasing numbers of terms $k$ in the expansion  (\ref{eq:partition19.2}).
The calculations were performed for $T=157$ MeV and $V=32$ fm$^3$, values
that are relevant to freezeout conditions in e$^+$e$^-$ annihilation. It
is clear from  this figure that the  Boltzmann approximation is by far not
sufficient to reproduce the quantum statistics results. The pion
yield under Boltzmann approximation deviates by more than $7\%$
from the  exact quantum statistics result. For kaons this
difference is only $1\%$.
While for kaons such an error is comparable to the 1 $\sigma$ error of 
the data in e$^+$e$^-$ collisions, for pions the deviation is significantly 
larger than the error in the data \cite{pdg}. This underlines the importance of 
using quantum statistics for the calculation of multiplicities of light mesons. 
For pions, several terms are needed in the expansion (\ref{last}) to achieve a 
precision well below 1\%.
It is also clear from Fig.~\ref{fig:dev1} that the deviations depend on the mass of
the particles and decrease quickly with increasing mass.
For protons, the lightest fermions, the corresponding deviation from Fermi-Dirac 
statistics is below 0.1\% already with only the 
first term, substantiating the applicability of the Boltzmann approximation 
for all fermions.

\begin{figure}[htb]
\begin{center}
\includegraphics[width=0.62\textwidth]{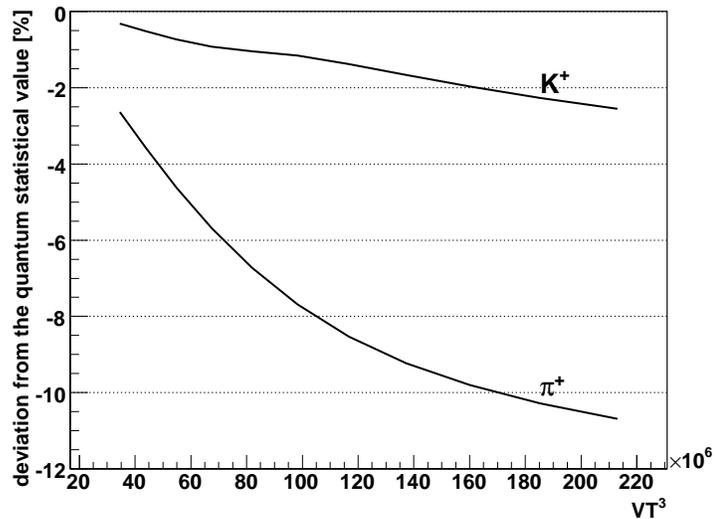}
\caption{Relative deviations of the pion and kaon multiplicities calculated  
using Boltzmann statistics from their  quantum statistics values as a 
function of $VT^3$.}
\label{fig:dev2}
\end{center}
\end{figure}

In Fig.~\ref{fig:dev2} we illustrate the relative error
of the calculated multiplicity of pions and kaons using the partition function
in the Boltzmann approximation  as a function of $VT^3$. Deviations from the
exact quantum statistics values are seen to increase with $VT^3$. 
This is due to the contribution of higher order terms in 
the expansion (\ref{last}), which are suppressed for small $VT^3$. 
For light bosons the largest deviations from the quantum statistics appear 
for large values of $VT^3$ i.e. when the system approaches GC thermodynamics.

\section{Conclusions}

We have presented a method to calculate the canonical partition function 
for the hadron resonance gas that accounts for exact  conservation of 
baryon number, electric change, strangeness, charmness  and bottomness. 
We have taken into account quantum statistics for light bosons and applied  
the Boltzmann approximation for fermions and heavier bosons. 
The  results obtained here  are an extension of previous studies which  
were restricted to the conservation of only three quantum numbers within 
the Boltzmann approximation \cite{Cleymans:1997ib}.
Our analytical expression for the partition function, which is represented 
as a series of  Bessel functions,  is stable in numerical implementations. 
It can be used for any value of the initial quantum numbers  of  the system 
and  for arbitrary thermal parameters. As an application of our results 
we have discussed the importance of quantum statistics in the calculations 
of pion and kaon multiplicities.
The canonical partition function which we have derived can be used to analyze
different thermodynamical properties of the hadron resonance gas with the
constraint of exact quantum number conservation. It can be also used 
to describe particle production in elementary and in heavy ion collisions 
within the statistical thermal model.

Acknowledgements:  
We acknowledge the support of the Alliance Program of the Helmholtz 
Association HA216/EMMI.
K.R. acknowledges partial support from the Polish Ministry of Science and 
Higher Education (MENiSW) and the Deutsche Forschungsgemeinschaft (DFG) under 
the Mercator Programme.

\end{document}